\documentclass[12 pt,twoside,aps,prd,amsmath,amssymb,
tightenlines,showpacs,showkeys,eqsecnum]{revtex4-1}

\usepackage{graphicx}
\usepackage{mathptmx}
\usepackage{bm}
{\hspace*{\fill}{\protect\small {\bf Bijan~Saha}} \hspace*{\fill} }
{\hspace*{\fill} {\protect\small {\bf Spinor field in Bianchi
type-$IX$ space-time}} \hspace*{\fill} } 

\newcommand {\cG}{\cal G}
\newcommand {\cD}{\cal D}
\newcommand {\bg}{\bar \gamma}
\newcommand {\G}{\Gamma}
\newcommand {\bp}{\bar \psi}

\def\myfigure #1#2#3#4
{\begin{figure}[ht]\begin{center}
\includegraphics[width=#2 \textwidth]{#1.eps}
\parbox[t]{#4\textwidth}{\caption{#3}\label{#1}}
\end{center}\end{figure}}

\def \myfigures #1#2#3#4#5#6#7#8
{\begin{figure}[ht]
    \begin{center}
        \includegraphics[width=#2 \textwidth]{#1.eps}
        \hfill
        \includegraphics[width=#6 \textwidth]{#5.eps}
        \parbox[t]{#4\textwidth}{\caption {#3}\label{#1}}
        \hfill
        \parbox[t]{#8\textwidth}{\caption {#7}\label{#5}}
    \end{center}
\end{figure} }

\begin{document}
\baselineskip -24pt
\title{Spinor field in Bianchi type-$IX$ space-time}
\author{Bijan Saha}
\affiliation{Laboratory of Information Technologies\\
Joint Institute for Nuclear Research\\
141980 Dubna, Moscow region, Russia} \email{bijan@jinr.ru}
\homepage{http://spinor.bijansaha.ru}

\hskip 1 cm

\begin{abstract}
Within the scope of Bianchi type-$IX$ we have studied the role of
spinor field in the evolution of the Universe. It is found that
unlike the diagonal Bianchi models in this case the components of
energy-momentum tensor of spinor field along the principal axis are
not the same, i.e. $T_1^1 \ne T_2^2 \ne T_3^3$, even in absence of
spinor field nonlinearity. The presence of nontrivial non-diagonal
components of energy-momentum tensor of the spinor field imposes
severe restrictions both on geometry of space-time and on the spinor
field itself. As a result the space-time turns out to be either
locally rotationally symmetric or isotropic. In this paper we
considered the Bianchi type-$IX$ space-time both for a trivial $b$,
that corresponds to standard $BIX$ and the one with a non-trivial
$b$.  It was found that a positive $\lambda_1$ gives rise to an
oscillatory mode of expansion, while a trivial $\lambda_1$ leads to
rapid expansion at the early stage of evolution.
\end{abstract}

\keywords{Spinor field, dark energy, anisotropic cosmological
models, isotropization}

\pacs{98.80.Cq}

\maketitle

\bigskip

\section{Introduction}

Nonlinear spinor fields plays a significant role in explaining the
evolution of the Universe at different its stages. It was shown by a
number of authors that the introduction of nonlinear spinor field
into the system can (i) give rise to a singularity-free Universe;
(ii) accelerate the isotropization process of initially anisotropic
space-time and (iii) generate late time acceleration of space-time
expansion
\cite{Saha1997GRG,Saha1997JMP,Saha2001PRD,Saha2004aPRD,Saha2004bPRD,PopPLB,
PopPRD,PopGREG,FabIJTP,Saha2006ECAA,kremer1,Saha2006PRD,Saha2006GnC,Saha2007RRP,
Saha2009aECAA,ELKO,FabGRG}. Moreover, it can simulate different
types of dark energy and perfect fluid
\cite{Krechet,Saha2010CEJP,Saha2010RRP,Saha2011APSS,Saha2012IJTP}.
Recently it was also found that the presence of non-diagonal
components of energy-momentum tensor of the spinor field imposes
severe restrictions to the space-time geometry as well
\cite{Saha2015CJP,Saha2015CnJP,Saha2016IJTP,Saha2016EPJP}.

In this paper we plan to extend our previous study to Bianchi
type-$IX$ cosmological model. One of the reasons to consider this
model is familiar solutions like the FRW Universe with positive
curvature, the de-sitter Universe , the Taub-Nut solutions etc. are
of Bianchi type-IX space-times. It should be noted that due to its
importance many authors have studied the evolution of the Universe
within the scope of a Bianchi type-$IX$ model. Bali {\it et. al.}
have studied the Bianchi type-$IX$ string cosmological models filled
with bulk viscous fluid \cite{bali2001,bali2003}, whereas such a
model for perfect fluid was investigated by Tyagi {\it et. al.} in
\cite{tyagi}. Analogous system with a time varying $\Lambda$-term
was studied in \cite{pradhan}. A scalar tensor theory of gravitation
within the framework of Bianchi type-$IX$ was studied by Reddy and
Naidu \cite{reddi}.

\section{Basic equation}

Let us consider the spinor field Lagrangian in the form

\begin{equation}
L = \frac{\imath}{2} \biggl[\bp \gamma^{\mu} \nabla_{\mu} \psi-
\nabla_{\mu} \bar \psi \gamma^{\mu} \psi \biggr] - m_{\rm sp} \bp
\psi - F, \label{lspin}
\end{equation}
where the nonlinear term $F$ describes the self-interaction of a
spinor field and can be presented as some arbitrary functions of
invariants generated from the real bilinear forms of a spinor field.
We consider the case when $F = F(K)$ with $K$ taking one of the
followings values $\{I,\,J,\,I + J,\,I - J\}$. By virtue of Fierz
theorem this comes out to be the most general form of spinor field
nonlinearity.

Here $\nabla_{\mu}$ covariant derivative of the spinor field having
the form

\begin{subequations}
\begin{eqnarray}
\nabla_{\mu} \psi &=& \partial_\mu \psi - \G_\mu \psi,\\
\nabla_{\mu} \bp &=& \partial_\mu \bp + \bp \G_\mu,
\end{eqnarray}
\end{subequations}
where $\G_\mu$ is the spinor affine connection defined as

\begin{equation}
\Gamma_\mu = \frac{1}{4} \bg_{a} \gamma^\nu \partial_\mu e^{(a)}_\nu
- \frac{1}{4} \gamma_\rho \gamma^\nu \Gamma^{\rho}_{\mu\nu},
\label{sfc}
\end{equation}
where $\bg_a$ are the Dirac matrices in flat space-time,
$\gamma_\nu$ are the Dirac matrices in curved space-time,
$e^{(a)}_\nu$ are the tetrad and $\Gamma^{\rho}_{\mu\nu}$ are the
Christoffel symbols.

The energy momentum tensor of the spinor field is given by
\begin{eqnarray}
T_{\mu}^{\,\,\,\rho}&=& \frac{\imath}{4} g^{\rho\nu} \biggl(\bp
\gamma_\mu \nabla_\nu \psi + \bp \gamma_\nu \nabla_\mu \psi -
\nabla_\mu \bar \psi \gamma_\nu \psi - \nabla_\nu \bp \gamma_\mu
\psi \biggr) \,- \delta_{\mu}^{\rho} L \nonumber\\
&=&\frac{\imath}{4} g^{\rho\nu} \biggl(\bp \gamma_\mu
\partial_\nu \psi + \bp \gamma_\nu \partial_\mu \psi -
\partial_\mu \bar \psi \gamma_\nu \psi - \partial_\nu \bp \gamma_\mu
\psi \biggr)\nonumber\\
& - &\frac{\imath}{4} g^{\rho\nu} \bp \biggl(\gamma_\mu \G_\nu +
\G_\nu \gamma_\mu + \gamma_\nu \G_\mu + \G_\mu \gamma_\nu\biggr)\psi
 \,- \delta_{\mu}^{\rho} L. \label{temsp0}
\end{eqnarray}

Bianchi type $IX$ space-time ($BIX$) is given by

\begin{equation}
ds^2 =  dt^2 - a_1^2(t) dx_1^2 - [h^2(x_3) a_1^2(t) + f^2(x_3)
a_2^2(t)] dx_2^2 -  a_3^2 dx_3^2 + 2 a_1^2(t)h(x_3)dx_1 dx_2,
\label{bii-ix}
\end{equation}
with $a_1,\,a_2,\,a_3$ being the functions of $t$. Here $h$ and $f$
are the functions of $x_3$. Here we consider the Bianchi type $IX$
space-time, which imposes the following restriction of $f$, namely
\begin{equation}
\delta = - \frac{f^{\prime\prime}}{f} = 1 \Rightarrow
f^{\prime\prime} + f = 0. \label{BIX}
\end{equation}
It should be noted that it is customary to assume $f (x_3) = \sin
(x_3)$ and  $h (x_3) = \cos (x_3)$. We don't write these concrete
expressions for the functions $f (x_3)$ and $h (x_3)$ here, but do
it later in due course.

To find the spinor affine connection \eqref{sfc} we have to know the
tetrad corresponding to the metric \eqref{bii-ix}. Exploiting the
well known relation
\begin{equation}
g_{\mu \nu} = e_\mu^{(a)} e_\nu^{(b)} \eta_{ab}, \label{tetmet}
\end{equation}

we choose the tetrad corresponding to \eqref{bii-ix} as follows:

\begin{equation}
e_0^{(0)} = 1, \quad e_1^{(1)} = a_1, \quad e_2^{(2)} = a_2 f, \quad
e_3^{(3)} = a_3, \quad e_2^{(1)} = -a_1 h. \label{tetradii-ix}
\end{equation}

From
\begin{equation}
\gamma_{\mu} = e_\mu^{(a)} \bg_a, \quad \gamma^\nu = e^\nu_{(b)}
\bg^b, \label{gamgg}
\end{equation}
such that
\begin{equation}
e_\mu^{(a)} e^\mu_{(b)} = \delta^a_b, \quad e_\mu^{(a)} e^\nu_{(a)}
=  \delta^\nu_\mu, \label{tetinv}
\end{equation}

one now finds
\begin{eqnarray}
\gamma_0 &=& \bg_0 = \bg^0, \quad  \gamma_1 =  a_1 \bg_1 = - a_1
\bg^1, \nonumber \\ \gamma_2 &=& - a_1 h \bg_1 + a_2 f \bg_2 = a_1 h
\bg^1 - a_2 f \bg^2,  \quad \gamma_3 = a_3 \bg_3 = - a_3 \bg^3,
\label{gbgii-ix}
\end{eqnarray}
where we take into account that
$$\bg^0 = \bg_0, \quad \bg^1 = -\bg_1, \quad \bg^2 = -\bg_2, \quad
\bg^3 = -\bg_3.$$

Using the laws of raising and lowering the indices one also finds
\begin{equation}
\gamma^0 = \bg^0, \quad  \gamma^1 = \frac{1}{a_1} \bg^1 +
\frac{h}{a_2 f} \bg^2, \quad \gamma^2 = \frac{1}{a_2 f} \bg^2, \quad
\gamma^3 = \frac{1}{a_3} \bg^3. \label{gbgii-ixup}
\end{equation}

Now we are ready to compute spinor affine connections using
\eqref{sfc}:

\begin{subequations}
\begin{eqnarray}
\G_1 &=&  \frac{1}{2} \dot a_1 \bg^1\bg^0 - \frac{1}{4} \frac{a_1^2
h^\prime}{ a_2 a_3 f} \bg^2\bg^3, \label{G1ii-ix}\\
\G_2 &=&   \frac{1}{2} f \dot a_2 \bg^2\bg^0 - \frac{1}{2} h \dot
a_1 \bg^1\bg^0 - \frac{1}{4}\frac{a_1 h^\prime}{a_3}\bg^1\bg^3 +
\frac{1}{2} \frac{a_2 f^\prime}{ a_3}  \bg^2\bg^3 +
\frac{1}{4}\frac{a_1^2 h h^\prime}{a_2 a_3 f}\bg^2\bg^3,
\label{G2ii-ix}\\
\G_3 &= &  \frac{1}{2} \dot a_3 \bg^3\bg^0 + \frac{1}{4} \frac{a_1
h^\prime}{a_2 f} \bg^1 \bg^2, \label{G3ii-ix}\\
\G_0 &= &  0.\label{G0ii-ix}
\end{eqnarray}
\end{subequations}

Then one finds
\begin{subequations}
\begin{eqnarray}
\gamma^\mu \G_\mu &=&   - \frac{1}{2} \frac{\dot V}{V}\bg^0 -
\frac{\imath}{4}\frac{ a_1 h^\prime}{ a_2 a_3 f}\bg^5\bg^0 -
\frac{1}{2}\frac{f^\prime}{ a_3 f} \bg^3, \label{gGii-ix}\\
\G_\mu \gamma^\mu &=&  \frac{1}{2} \frac{\dot V}{V}\bg^0 -
\frac{\imath}{4}\frac{a_1 h^\prime}{a_2 a_3 f}\bg^5\bg^0 +
\frac{1}{2}\frac{f^\prime}{a_3 f} \bg^3, \label{Ggii-ix}
\end{eqnarray}
\end{subequations}
where we introduce the volume scale
\begin{equation}
V = a_1 a_2 a_3 \label{Vdef-II-IX}
\end{equation}
and $\bg^5 = - \imath \bg^0 \bg^1 \bg^2 \bg^3.$

The spinor field equations corresponding to the Lagrangian
\eqref{lspin} has the form

\begin{subequations}
\label{speq}
\begin{eqnarray}
\imath\gamma^\mu \nabla_\mu \psi - m_{\rm sp} \psi - {\cD}
\psi -  \imath {\cG} \gamma^5 \psi &=&0, \label{speq1} \\
\imath \nabla_\mu \bp \gamma^\mu +  m_{\rm sp} \bp + {\cD} \bp +
\imath {\cG}  \bp \gamma^5 &=& 0, \label{speq2}
\end{eqnarray}
\end{subequations}
where ${\cD} = 2 S F_K K_I$ and ${\cG} = 2 P F_K K_J$.

In view of \eqref{gGii-ix} and \eqref{Ggii-ix} the system
\eqref{speq} can be rewritten as
\begin{subequations}
\label{speqIIcom}
\begin{eqnarray}
\imath\bg^0 \dot \psi + \frac{\imath}{2}  \frac{\dot V}{V} \bg^0
\psi + \frac{1}{4} \frac{a_1 h^\prime}{ a_2 a_3 f}\bg^5 \bg^0 \psi +
\frac{\imath}{2} \frac{f^\prime}{ a_3 f} \bg^3 \psi - \left[m_{\rm
sp} +{\cD}\right]
\psi -  \imath {\cG} \bg^5 \psi &=&0, \label{speq1IIcom} \\
\imath \dot \bp \bg^0 + \frac{\imath}{2} \frac{\dot V}{V} \bp \bg^0
-\frac{1}{4}\frac{a_1 h^\prime}{ a_2 a_3 f} \bp \bg^5 \bg^0 +
\frac{\imath}{2} \frac{f^\prime}{ a_3 f} \bp \bg^3 + \left[m_{\rm
sp} + {\cD}\right] \bp +  \imath {\cG}  \bp \bg^5 &=& 0.
\label{speq2IIcom}
\end{eqnarray}
\end{subequations}

In view of \eqref{speq} the spinor field Lagrangian can be written
as

\begin{eqnarray}
L & = & \frac{\imath}{2} \bigl[\bp \gamma^{\mu} \nabla_{\mu} \psi-
\nabla_{\mu} \bar \psi \gamma^{\mu} \psi \bigr] - m_{\rm sp} \bp
\psi - F(K)
\nonumber \\
& = & \frac{\imath}{2} \bp [\gamma^{\mu} \nabla_{\mu} \psi - m_{\rm
sp} \psi] - \frac{\imath}{2}[\nabla_{\mu} \bar \psi \gamma^{\mu} +
m_{\rm sp} \bp] \psi
- F(K),\nonumber \\
& = & 2 (I F_I + J F_J) - F = 2 K F_K - F(K), \label{lspin01II}
\end{eqnarray}
from \eqref{temsp0} one finds the following components of
energy-momentum tensor

\begin{subequations}
\label{TEVBII_IX}
\begin{eqnarray}
T^{0}_{0} & = &   m_{\rm sp} S + F(K) -  \frac{1}{4} \frac{a_1
h^\prime}{a_2 a_3 f}A^0, \label{ApT00f}\\
T^{1}_{1} & = & \left[ F(K) - 2 K F_K\right] +
\frac{1}{4}\frac{h}{a_2^2 f} \left(a_1 \dot a_2 - \dot a_1
a_2\right) A^3 + \frac{1}{4}\frac{a_1 h}{a_2 a_3
f}\left(\frac{f^\prime}{f} - \frac{h^\prime}{h}\right)
A^0, \label{ApT11f}\\
T^{2}_{2} & = & \left[ F(K) - 2 K F_K\right] -
\frac{1}{4}\frac{h}{a_2^2 f} \left(a_1 \dot a_2 - \dot a_1
a_2\right) A^3 - \frac{1}{4}\frac{a_1 h}{a_2 a_3
f}\left(\frac{f^\prime}{f} - \frac{h^\prime}{h}\right) A^0,
\label{ApT22f}\\
T^{3}_{3} & = & \left[ F(K) - 2 K F_K\right] + \frac{1}{4}\frac{a_1
h}{a_2 a_3 f} \frac{h^\prime}{h} A^0, \label{ApT33f}\\
T^{1}_{2} & = & \frac{1}{4} \left(\frac{f}{a_1^2} - \frac{h^2}{a_2^2
f}\right) \left( a_1 \dot a_2 - \dot a_1 a_2\right) A^3 +
\left[\frac{a_1 h^2}{a_2 a_3 f} \left(\frac{1}{2}\frac{h^\prime}{h}
- \frac{1}{4}\frac{f^\prime}{f}\right) + \frac{1}{4}\frac{a_2 f}{a_1
a_3}\frac{f^\prime}{f}\right] A^0,\label{ApT12f}\\
T^{2}_{1} & = & \frac{1}{4}\frac{1}{a_2^2 f} \left( a_1 \dot a_2 -
\dot a_1 a_2\right) A^3 + \frac{1}{4}\frac{a_1}{a_2 a_3
f}\frac{f^\prime}{f} A^0,\label{ApT21f}\\
T^{2}_{3} & = & - \frac{1}{4}\frac{1}{a_2^2 f} \left(\dot a_2 a_3 -
a_2 \dot a_3\right) A^1,  \label{ApT23f}\\
T^{3}_{2} & = & - \frac{1}{4} \frac{1}{a_3^2} \left[ h\left(\dot a_1
a_3 - a_1 \dot a_3\right)A^2 + f \left(\dot a_2 a_3 - a_2 \dot
a_3\right)A^1 \right],\label{ApT32f}\\
T^{1}_{3} & = & \frac{1}{4}\frac{1}{a_1^2 }  \left(  \dot a_1 a_3 -
a_1 \dot a_3 \right) A^2 - \frac{1}{4} \frac{h}{a_2^2 f} \left(\dot
a_2 a_3 - a_2 \dot a_3\right) A^1,  \label{ApT13f}\\
T^{3}_{1} & = & \frac{1}{4} \frac{1}{a_3^2} \left( \dot a_1 a_3 -
a_1 \dot a_3 \right) A^2, \label{ApT31f}\\
T^{0}_{1} & = & \frac{1}{4} \left[ \frac{a_1^2 h^\prime}{ a_2 a_3
f} A^1 +  \frac{a_1 f^\prime}{a_3}  A^2\right], \label{ApT01f}\\
T^{1}_{0} & = & \frac{1}{4}\left[\left( \frac{a_1 h f^\prime}{a_2^2
a_3 f^2} + \frac{h f^\prime}{a_2 a_3 f} - \frac{h^\prime}{a_2 a_3
f}\right) A^1 -  \frac{f^\prime}{a_1 a_3 }  A^2\right],
\label{ApT10f}\\
T^{0}_{2} & = & - \frac{1}{4}\left[\left(\frac{a_1 f^\prime}{a_3} +
\frac{a_2 f f^\prime}{a_3} +  \frac{a_1^2 h h^\prime}{ a_2 a_3 f}
\right) A^1 + \frac{a_1 h f^\prime}{a_3}\,A^2\right],
\label{ApT02f}\\
T^{2}_{0} & = & \frac{1}{4}\left(\frac{f\prime}{a_2 a_3 f} +
\frac{a_1 f\prime}{a_2^2 a_3 f^2}\right) A^1,  \label{ApT20f}\\
T^{0}_{3} & = & 0, \label{ApT03f}\\
T^{3}_{0} & = & 0. \label{ApT30f}
\end{eqnarray}
\end{subequations}

From \eqref{TEVBII_IX} we see that the spinor field distribution
along the main axis is anisotropic, i.e. $T_1^1 \ne T_2^2 \ne T_3^3$
and these components do not vanish even in absence of spinor field
nonlinearity.

The components of Einstein tensor corresponding to \eqref{bii-ix}
are

\begin{subequations}
\label{EEBII_IX}
\begin{eqnarray}
G^1_1 &=& - \left(\frac{\ddot a_2}{a_2} + \frac{\ddot a_3}{a_3} +
\frac{\dot a_2}{a_2}\frac{\dot a_3}{a_3}\right) + \frac{a_1^2
h^2}{a_2^2 a_3^2 f^2}\left(\frac{3}{4}\frac{h^{\prime 2}}{h^2}
 + \frac{1}{2}\frac{h^{\prime \prime}}{h}
 - \frac{1}{2}\frac{h^\prime}{h}\frac{f^\prime}{f}\right) + \frac{1}{a_3^2}\frac{f^{\prime \prime}}{f},\label{EE11}\\
 G^2_1 &=& \frac{1}{2} \frac{a_1^2 h}{a_2^2 a_3^2
 f^2} \left(\frac{h^{\prime \prime}}{h} - \frac{h^\prime}{h}\frac{f^\prime}{f}\right),
 \label{EE12}\\
 G^1_2 &=& h \left(\frac{\ddot a_2}{a_2} - \frac{\ddot a_1}{a_1} +
\frac{\dot a_2}{a_2}\frac{\dot a_3}{a_3} - \frac{\dot
a_1}{a_1}\frac{\dot a_3}{a_3}\right) +
\frac{h}{a_3^2}\left(\frac{1}{2}\frac{h^{\prime\prime}}{h} -
\frac{f^{\prime\prime}}{f}  - \frac{1}{2}
\frac{h^\prime}{h}\frac{f^\prime}{f}\right) \nonumber\\ & &  +
\frac{a_1^2 h^3 }{a_2^2 a_3^2 f^2} \left(\frac{1}{2}
\frac{h^\prime}{h}\frac{f^\prime}{f}  - \frac{1}{2}\frac{
h^{\prime\prime}}{h} - \frac{h^{\prime 2}}{h^2}\right),
\label{EE21}\\
G_2^2 &=& - \left(\frac{\ddot a_3}{a_3} + \frac{\ddot a_1}{a_1} +
\frac{\dot a_3}{a_3}\frac{\dot a_1}{a_1}\right) - \frac{1}{2}
\frac{a_1^2 h^2}{a_2^2 a_3^2 f^2} \left(\frac{ h^{\prime\prime}}{h}
- \frac{h^\prime}{h}\frac{f^\prime}{f} + \frac{1}{2}
\frac{h^{\prime 2}}{h^2}\right), \label{EE22}\\
G^3_3 &=& - \left(\frac{\ddot a_1}{a_1} + \frac{\ddot a_2}{a_2} +
\frac{\dot a_1}{a_1}\frac{\dot a_2}{a_2}\right) -
\frac{1}{4}\frac{a_1^2 h^2}{a_2^2 a_3^2 f^2}
\frac{ h^{\prime 2}}{h^2}, \label{EE33}\\
G^0_3 &=& \left(\frac{\dot a_2}{a_2} - \frac{\dot a_3}{a_3}\right) \frac{f^\prime}{f}, \label{EE30}\\
G^3_0 &=& - \frac{1}{a_3^2}\left(\frac{\dot a_2}{a_2} - \frac{\dot a_3}{a_3}\right) \frac{f^\prime}{f}, \label{EE03}\\
G^0_0 &=& -\left(\frac{\dot a_1}{a_1}\frac{\dot a_2}{a_2} +
\frac{\dot a_2}{a_2}\frac{\dot a_3}{a_3} + \frac{\dot
a_3}{a_3}\frac{\dot a_1}{a_1} \right) + \frac{1}{4}\frac{a_1^2
 h^2}{a_2^2 a_3^2 f^2}\frac{h^{\prime 2}}{h^2} + \frac{1}{a_3^2}\frac{f^{\prime\prime}}{
f}.\label{EE00}
\end{eqnarray}
\end{subequations}
From \eqref{EEBII_IX} one finds the following relations between its
components:
\begin{equation}
G^1_2 = h \left(G^2_2 - G^1_1\right) + \left(h^2 + \frac{a_2^2
f^2}{a_1^2}\right) G^2_1. \label{G2112}
\end{equation}

From \eqref{TEVBII_IX} it can be shown that

\begin{equation}
T^1_2 = h \left(T^2_2 - T^1_1\right) + \left(h^2 + \frac{a_2^2
f^2}{a_1^2}\right) T^2_1. \label{T2112}
\end{equation}

Moreover from \eqref{TEVBII_IX} it follows that
\begin{equation}
T^3_2 = \frac{a_2^2 f^2}{a_3^2} T^2_3 - h T^3_1, \label{T322331}
\end{equation}

\begin{equation}
T^1_3 =  h T^2_3  + \frac{a_3^2}{a_1^2} T^3_1, \label{T132331}
\end{equation}

\begin{equation}
T^1_0 =  h T^2_0  - \frac{1}{a_1^2} T^0_1, \label{T102001}
\end{equation}

and

\begin{equation}
T^0_2 = - a_2^2 f^2 T^2_0  - h  T^0_1. \label{T022001}
\end{equation}

Then the system of Einstein equations

\begin{equation}
G^\mu_\nu = - \kappa T^\mu_\nu, \label{EE}
\end{equation}

on account of linearly dependent components takes the form
\begin{subequations}
\label{EECom}
\begin{eqnarray}
\left(\frac{\ddot a_2}{a_2} + \frac{\ddot a_3}{a_3} + \frac{\dot
a_2}{a_2}\frac{\dot a_3}{a_3}\right) &-&\frac{1}{2} \frac{a_1^2
h^2}{a_2^2 a_3^2 f^2}\left(\frac{h^{\prime \prime}}{h}
 - \frac{h^\prime}{h}\frac{f^\prime}{f} + \frac{3}{2}\frac{h^{\prime 2}}{h^2}\right)
 - \frac{1}{a_3^2}\frac{f^{\prime \prime}}{f} \label{EECom11}
 \\
 &=&\kappa \left[ \left( F(K) - 2 K F_K\right) +
\frac{1}{4}\frac{a_1 h}{a_2 f} \left(\frac{\dot a_2}{a_2} - \frac{
\dot a_1}{ a_1}\right) A^3 + \frac{1}{4}\frac{a_1 h}{a_2 a_3
f}\left(\frac{f^\prime}{f} - \frac{h^\prime}{h}\right) A^0\right],
  \nonumber\\
\left(\frac{\ddot a_3}{a_3} + \frac{\ddot a_1}{a_1} + \frac{\dot
a_3}{a_3}\frac{\dot a_1}{a_1}\right) & + &\frac{1}{2} \frac{a_1^2
h^2}{a_2^2 a_3^2 f^2} \left(\frac{ h^{\prime\prime}}{h} -
\frac{h^\prime}{h}\frac{f^\prime}{f} + \frac{1}{2} \frac{h^{\prime
2}}{h^2}\right) \label{EECom22}\\
&=& \kappa \left[ \left(F(K) - 2 K F_K\right) - \frac{1}{4}\frac{a_1
h}{a_2 f} \left(\frac{ \dot a_2}{a_2} - \frac{\dot a_1}{ a_1}\right)
A^3 - \frac{1}{4}\frac{a_1 h}{a_2 a_3 f}\left(\frac{f^\prime}{f} -
\frac{h^\prime}{h}\right) A^0\right],   \nonumber \\
 \left(\frac{\ddot a_1}{a_1} + \frac{\ddot a_2}{a_2} +
\frac{\dot a_1}{a_1}\frac{\dot a_2}{a_2}\right) & + &
\frac{1}{4}\frac{a_1^2 h^{\prime 2}}{a_2^2 a_3^2 f^2}  = \kappa
\left[ \left(F(K) - 2 K F_K\right) + \frac{1}{4}\frac{a_1
h^\prime}{a_2 a_3 f}  A^0\right], \label{EECom33}\\
\left(\frac{\dot a_1}{a_1}\frac{\dot a_2}{a_2} + \frac{\dot
a_2}{a_2}\frac{\dot a_3}{a_3} + \frac{\dot a_3}{a_3}\frac{\dot
a_1}{a_1} \right) & - & \frac{1}{4}\frac{a_1^2
 h^{\prime 2}}{a_2^2 a_3^2 f^2} -
\frac{1}{a_3^2}\frac{f^{\prime\prime}}{ f} =\kappa \left[ m_{\rm sp}
S + F(K) -  \frac{1}{4} \frac{a_1
h^\prime}{a_2 a_3 f}A^0\right], \label{EECom00}\\
\frac{1}{2} \frac{a_1^2 h}{a_2^2 a_3^2
 f^2} \left(\frac{h^{\prime \prime}}{h} -
 \frac{h^\prime}{h}\frac{f^\prime}{f}\right) &=& -\kappa \frac{1}{4}
 \frac{1}{a_2 f}   \left[\left( \frac{ \dot a_2}{a_2} -
\frac{\dot a_1}{ a_1}\right) A^3 + \frac{a_1}{ a_3
}\frac{f^\prime}{f} A^0\right], \label{EECom21}\\
\left(\frac{\dot a_2}{a_2} - \frac{\dot a_3}{a_3}\right)
\frac{f^\prime}{f} &=& 0, \label{EECom03}\\
0 & =&\left(\frac{\dot a_2}{a_2} -
\frac{\dot a_3}{a_3}\right) A^1, \label{EECom23}\\
0 & =&  \left( \frac{\dot a_1}{a_1} -
\frac{\dot a_3}{a_3} \right) A^2, \label{EECom31}\\
0 & =&  \left[ \frac{a_1 h^\prime}{ a_2
f} A^1 +   f^\prime  A^2\right], \label{EECom01}\\
0 & =& \left(1 + \frac{a_1}{a_2 f}\right) \frac{f^\prime}{f}\, A^1.
\label{EECom20}
\end{eqnarray}
\end{subequations}

 Then in view of $\frac{f^\prime}{f}
\ne 0$ from \eqref{EECom03} we find $\left(\frac{\dot a_2}{a_2} -
\frac{\dot a_3}{a_3}\right) = 0$. On the other hand for same reason
\eqref{EECom20} yields $A^1 = 0$, whereas inserting $A^1 = 0 $ into
\eqref{EECom01} we obtain $A^2 = 0$. Thus in this case from
\eqref{EECom03} - \eqref{EECom20} we have

\begin{eqnarray}
A^1 = 0, \quad A^2 = 0, \quad \left(\frac{\dot a_2}{a_2} -
\frac{\dot a_3}{a_3}\right) = 0. \label{03-20}
\end{eqnarray}
In view of $A^2 = 0$ the equation \eqref{EECom31} yields two
possibilities:

\begin{eqnarray}
\left(\frac{\dot a_1}{a_1} - \frac{\dot a_3}{a_3}\right) \ne 0,
\label{3-1ne0}
\end{eqnarray}
which means the model is rotationally symmetric, or
\begin{eqnarray}
\left(\frac{\dot a_1}{a_1} - \frac{\dot a_3}{a_3}\right) = 0,
\label{3-1eq0}
\end{eqnarray}
which means the model is isotropic. Note that in case of diagonal
energy-momentum tensor we have strictly locally rotationally
symmetric space-time \cite{SahaGnC2013}.

It should be noted that for the diagonal Bianchi models the volume
scale $V$ plays crucial role in the evolution of the Universe. $V$
plays important role for non-diagonal Bianchi models too. So let us
now write the euqtion for $V$. Summation of \eqref{EECom11},
\eqref{EECom22}, \eqref{EECom33} and 3 times \eqref{EECom00} gives

\begin{eqnarray}
\frac{\ddot V}{V} - \frac{1}{4} \frac{a_1^2
 }{a_2^2 a_3^2}\frac{h^{\prime 2}}{f^2} -
 \frac{1}{2a_3^2}\frac{f^{\prime\prime}}{f} = \frac{3
 \kappa}{2}\left[m_{\rm sp} S + 2 \left(F - K F_K\right) -
 \frac{1}{2}  \frac{a_1}{a_2 a_3}\frac{h^\prime}{f}A^0\right]. \label{EqV}
\end{eqnarray}

As one sees, to find the solution of \eqref{EqV} one need to know
$f$,\,$h$,\,$A_0$, the spinor field nonlinearity as well as the
metric functions, or at least their expressions in terms of $V$.

To find the metric functions in terms of $V$ we use the
proportionality condition. In doing so let us compute the expansion
and shear corresponding to metric \eqref{bii-ix}. Let the
four-velocity is given by $u^\mu = (1,\,0,\,0,\,0)$. Then for the
expansion we have
\begin{eqnarray}
\vartheta &=& u^\mu_{;\mu} = u^\mu_{,\mu} + \G^\mu_{\mu \alpha}
u^\alpha = \G^\mu_{\mu 0} \nonumber \\
&=&  \G^1_{10} + \G^2_{30} + \G^3_{30} = \frac{\dot a_1}{a_1}+
\frac{\dot a_2}{a_2} + \frac{\dot a_3}{a_3} = \frac{\dot V}{V}.
\label{Expan}
\end{eqnarray}
The shear is given by
\begin{eqnarray}
\sigma_{\alpha \beta} &=&  \frac{1}{2} \left(u_{\alpha; \nu}
P^{\nu}_\beta + u_{\beta; \nu} P^{\nu}_\alpha\right) - \frac{1}{3}
\vartheta P_{\alpha \beta}.\label{shearcomp}
\end{eqnarray}
Taking into account that $u_\mu = u^\nu g_{\nu \mu} = u^0 g_{0\mu} =
(1,\,0,\,0,\,0)$, \quad  $u_{\alpha;\nu} = u_{\alpha,\nu} -
\G^\mu_{\alpha\nu} u_{\mu} = - \G^0_{\alpha\nu}$, \quad $P_{\alpha
\beta} = g_{\alpha \beta} - u_\alpha u_\beta$ and $P^\alpha_\beta =
\delta^\alpha_\beta - u^\alpha u_\beta$ from \eqref{shearcomp} we
find
\begin{eqnarray}
\sigma_{\alpha \beta} &=&  \frac{1}{2} \left(-\G^0_{\alpha\nu}
P^{\nu}_\beta - \G^0_{\beta\nu} P^{\nu}_\alpha\right) - \frac{1}{3}
\vartheta P_{\alpha \beta}.\label{shearcomp1}
\end{eqnarray}

Using \eqref{shearcomp1} we find

\begin{subequations}
\label{sigmixf}
\begin{eqnarray}
\sigma_1^1 &=&  \frac{\dot a_1}{a_1} - \frac{1}{3} \vartheta,
\label{shearcomp111mixf}\\
\sigma_2^2 &=&   \frac{\dot a_2}{a_2} - \frac{1}{3} \vartheta, \label{shearcomp122mixf}\\
\sigma_3^3 &=&  \frac{\dot a_3}{a_3} - \frac{1}{3} \vartheta, \label{shearcomp133mixf}\\
\sigma_2^1 &=&  h \left(\frac{\dot a_2}{a_2} - \frac{\dot a_1}{a_1}\right), \label{shearcomp112mixf}\\
\sigma_1^2 &=&  = 0. \label{shearcomp121mixf}
\end{eqnarray}
\end{subequations}

Thus we see that the diagonal components of shear tensor
$\sigma_\alpha^\beta$ does not depend on $f$ and $h$.

Let us assume the proportionality condition

\begin{equation}
\sigma_1^1 = q_1 \vartheta, \quad q_1 = {\rm const.} \label{propcon}
\end{equation}
which after inserting \eqref{shearcomp111mixf} and \eqref{Expan}
gives
\begin{equation}
\frac{\dot a_1}{a_1}=\left(q_1 + \frac{1}{3}\right) \vartheta.
\label{propcon1}
\end{equation}
In view of $\vartheta = \dot V/V$ from \eqref{propcon1} in one hand
we find
\begin{equation}
\frac{\dot a_1}{a_1}=\left(q_1 + \frac{1}{3}\right) \frac{\dot
V}{V}, \label{a1V}
\end{equation}
with
\begin{equation}
a_1 = q_2 V^{q_1 + 1/3}, \quad q_2 = {\rm const.} \label{a1Vexp}
\end{equation}
On the other hand from \eqref{EECom03} one finds
\begin{equation}
a_2 = q_3 a_3, \quad q_3 = {\rm const.} \label{a2Vexp}
\end{equation}

Then from \eqref{Vdef-II-IX} one finds

\begin{equation}
a_3 = \frac{1}{\sqrt{q_2 q_3}} V^{1/3 - q_1/2}. \label{a3Vexp}
\end{equation}

So finally we can write the expressions for metric functions in
terms of $V$ as

\begin{equation}
a_1 = X_i V^{Y_i}, \quad \prod_{i = 1}^3 X_i  = 1, \quad \sum_{i =
1}^3 Y_i = 0.  \label{a123Vexp}
\end{equation}
In this concrete case we have $X_1 = q_2$,\,\,\,$X_2 =
\sqrt{{q_3}/{q_2}}$,\,\,\, $X_3 = {1}/{\sqrt{q_2 q_3}}$,\,\,\,$ Y_1
= q_1 + 1/3,$\,\, and \,\,$Y_2 = Y_3 = 1/3 - q_1/2.$

From \eqref{a123Vexp} one finds that ${a_1}/{(a_2a_3)} = q_2^2
V^{2q_1 - 1/3}$. Then the equation for $V$ \eqref{EqV}  can be
rewritten as

\begin{eqnarray}
\frac{\ddot V}{V} &-& \frac{1}{4}  q_2^4 V^{4q_1 - 2/3}
\left(\frac{h^\prime}{f}\right)^2-
 \frac{1}{2} q_2 q_3 V^{q_1 - 2/3} \frac{f^{\prime\prime}}{f}\nonumber\\ & =& \frac{3
 \kappa}{2}\left[m_{\rm sp} S + 2 \left(F - K F_K\right) -
 \frac{1}{2} q_2^2 V^{2q_1 - 1/3} \frac{h^\prime}{f} A^0\right], \label{EqVBIX}
\end{eqnarray}

From the spinor field equations \eqref{speq} we find that the
bilinear forms of the spinor field in this case obey the following
system of equations:

\begin{subequations}
\label{SpinInv}
\begin{eqnarray}
\dot S_0   + \frac{1}{2} \frac{a_1 h^\prime}{ a_2 a_3 f} P_0  + 4 P
F_K K_J A^0_0 & = & 0, \label{S0Inv}\\
\dot P_0   + \frac{1}{2} \frac{a_1 h^\prime}{ a_2 a_3 f} S_0  - 2
\left[m_{\rm sp} + 2 S F_K K_I\right] A^0_0 &=& 0, \label{P0Inv}\\
 \dot A^0_0   +
\frac{1}{2} \frac{f^\prime}{a_3 f} A^3_0  + 2 \left[m_{\rm sp} + 2 S
F_K K_I\right] P_0 + 4 P F_K K_J S_0 &=& 0, \label{A00Inv}\\
 \dot A^3_0   +
\frac{1}{2} \frac{f^\prime}{a_3 f} A^0_0  &=& 0. \label{A30Inv}
\end{eqnarray}
\end{subequations}
From \eqref{SpinInv} it can be easily shown that

\begin{equation}
P_0^2 - S_0^2 + \left(A_0^0\right)^2 - \left(A_0^3\right)^2  = {\rm
const.}, \label{InVCons}
\end{equation}
On the other hand from Fierz theorem we have

\begin{equation}
I_A = \left(A^0\right)^2 - \left(A^1\right)^2 - \left(A^2\right)^2 -
\left(A^3\right)^2 = -\left(S^2 + P^2\right), \label{InVCons1}
\end{equation}

Now taking into account that $A^1 = 0$ and $A^2 = 0$ from
\eqref{InVCons1} one finds

\begin{equation}
\left(A_0^0\right)^2 - \left(A_0^3\right)^2 = -\left(S_0^2 +
P_0^2\right), \label{InVCons2}
\end{equation}

Then inserting \eqref{InVCons2} into \eqref{InVCons} one finds
\begin{equation}
S = \frac{V_0}{V}, \quad V_0 = {\rm const.} \label{InVS0}
\end{equation}
It should be emphasized that in case of diagonal Bianchi space-time
we obtain the expression \eqref{InVS0} only when the spinor field
nonlinearity depends on $K = I = S^2$. Thus we see that in case of
$BIX$ space-time $S = V_0/V$ independent to our choice of $K$.

Let us now see, what happens if $K$ takes any of the following
expressions $\{J,\,\,I + J,\,\,I - J\}$. In case of diagonal Bianchi
models exact expressions were found for massless spinor field only.
So here we consider massless spinor field.  Then in case of $K = J$
from \eqref{P0Inv} we find

\begin{equation}
\dot P_0   + \frac{1}{2} q_2^2 V^{2q_1 - 1/3} \frac{h^\prime}{f} V_0
= 0. \label{P0Inv1}
\end{equation}
From \eqref{P0Inv1} one can formally express $K = J = P^2$  in terms
of $V$.

In case of $K = I + J$ we have
\begin{subequations}
\label{SpinInv2}
\begin{eqnarray}
\dot S_0   + \frac{1}{2} q_2^2 V^{2q_1 - 1/3} \frac{h^\prime}{f} P_0
+ 4 P F_K A^0_0 & = & 0, \label{S0Inv2}\\
\dot P_0   + \frac{1}{2} q_2^2 V^{2q_1 - 1/3} \frac{h^\prime}{f} S_0
- 4 S F_K  A^0_0 &=& 0. \label{P0Inv2}
\end{eqnarray}
\end{subequations}
Summation of \eqref{S0Inv2} multiplied by $S_0$ and \eqref{P0Inv2}
multiplied by $P_0$ gives
\begin{equation}
\left(S_0 \dot S_0 + P_0 \dot P_0\right)   +  q_2^2 V^{2q_1 - 1/3}
\frac{h^\prime}{f} S_0 P_0 = 0. \label{P0Inv21}
\end{equation}
Further in view of $S_0 = V_0$ from \eqref{P0Inv21} one finds
equation for $P_0$ analogous to \eqref{P0Inv1}. Knowing $J = P^2$ we
obtain the expression for $K = I + J$ in terms of $V$.

Finally, for $K = I - J$ we have
\begin{subequations}
\label{SpinInv3}
\begin{eqnarray}
\dot S_0   + \frac{1}{2} q_2^2 V^{2q_1 - 1/3} \frac{h^\prime}{f} P_0
- 4 P F_K A^0_0 & = & 0, \label{S0Inv3}\\
\dot P_0   + \frac{1}{2} q_2^2 V^{2q_1 - 1/3} \frac{h^\prime}{f} S_0
- 4 S F_K  A^0_0 &=& 0. \label{P0Inv3}
\end{eqnarray}
\end{subequations}
Subtraction of \eqref{P0Inv3} multiplied by $P_0$ from
\eqref{S0Inv3} multiplied by $S_0$ gives
\begin{equation}
\left(S_0 \dot S_0 - P_0 \dot P_0\right)  = 0, \label{P0inv22}
\end{equation}
with the solution
\begin{equation}
K = \left(I - J\right) = \left(S^2 - P^2\right)  =
\frac{V_0^2}{V^2}. \label{KImJ}
\end{equation}

Here it is interesting note that in absence of spinor field
nonlinearity the system \eqref{SpinInv} takes the form

\begin{subequations}
\label{SpinInvl}
\begin{eqnarray}
\dot S_0   + \frac{1}{2} \frac{a_1 h^\prime}{ a_2 a_3 f} P_0  & = & 0, \label{S0Invl}\\
\dot P_0   + \frac{1}{2} \frac{a_1 h^\prime}{ a_2 a_3 f} S_0  - 2
m_{\rm sp} A^0_0   &=& 0, \label{P0invl}\\
 \dot A^0_0   +
\frac{1}{2} \frac{f^\prime}{a_3 f} A^3_0  + 2 m_{\rm sp}
 P_0  &=& 0, \label{A00invl}\\
 \dot A^3_0   +
\frac{1}{2} \frac{f^\prime}{a_3 f} A^0_0  &=& 0. \label{A30invl}
\end{eqnarray}
\end{subequations}
One can easily find that this system too allows the first integral
\eqref{InVCons}. Moreover, in case of massless spinor field we find
\begin{equation}
P_0^2 - S_0^2 = {\rm const.}, \quad \left(A_0^0\right)^2 -
\left(A_0^3\right)^2  = {\rm const.}. \label{invSA}
\end{equation}
Now in view of \eqref{03-20} and \eqref{InVS0} Eq. \eqref{invSA} can
be rewritten as
\begin{equation}
P_0^2 - S_0^2 = {\rm const.}, \quad  -\left(P_0^2 + S_0^2\right) =
{\rm const.}, \label{invSA1}
\end{equation}
which gives
\begin{equation}
S_0 = {\rm const.} \quad  P_0 = {\rm const.} \label{SPC}
\end{equation}

Hence the behavior of invariants, constructed from bilinear spinor
forms categorically differs from that we obtain for diagonal Bianchi
models.

Let us now recall that the functions $f( x_3)$ can be concretize
using the restriction imposed on it, namely, from the \eqref{BIX}
one finds the following solutions for $f: f= \sin (x_3)$ and $f =
\cos (x_3)$. Following a number of authors we choose  $f = \sin
(x_3)$.  As far as  $h$ is concerned, it should be determined from
\eqref{EECom21}.  In doing so we go back to \eqref{EECom21} which
can be rewritten as
\begin{eqnarray}
 \frac{1}{f} \left(h^{\prime \prime} -
 \frac{f^\prime}{f} h^\prime\right) = -\frac{\kappa}{2}
 \frac{a_2 a_3^2}{a_1^2}   \left[\left( \frac{ \dot a_2}{a_2} -
\frac{\dot a_1}{ a_1}\right) A_0^3 + \frac{a_1}{ a_3
}\frac{f^\prime}{f} A_0^0\right], \label{EECom21n}
\end{eqnarray}

In view of \eqref{A30Inv}  this equations can be written as

\begin{eqnarray}
 \frac{1}{f} \left(h^{\prime \prime} -
 \frac{f^\prime}{f} h^\prime\right) = -\frac{\kappa}{2}
 \frac{a_2 a_3^2}{a_1^2}   \left[\left( \frac{ \dot a_2}{a_2} -
\frac{\dot a_1}{ a_1}\right) A_0^3 - 2 a_1 {\dot A_0^3}\right],
\label{EECom21n1}
\end{eqnarray}
Now the left hand side of \eqref{EECom21n1} depends of $x_3$ only,
while the right hand side depends on only $t$. So we can finally
write the following system

\begin{subequations}
\label{eqha3}
\begin{eqnarray}
\frac{1}{f} \left(h^{\prime \prime} -
 \frac{f^\prime}{f} h^\prime\right) &=& b, \label{eqh}\\
\frac{\kappa}{2}
 \frac{a_2a_3^2}{a_1^2}   \left[\left( \frac{ \dot a_2}{a_2} -
\frac{\dot a_1}{ a_1}\right) A_0^3 - 2 a_1 {\dot A_0^3}\right] &=& -
b, \label{eqa3}
\end{eqnarray}
\end{subequations}
From the foregoing equations in view of $f = \sin (x_3)$ and
\eqref{a123Vexp}  we finally obtain
\begin{subequations}
\begin{eqnarray}
h^{\prime \prime} +
 \cot (x_3) h^\prime  &=& b \sin (x_3), \label{eqh1}\\
\dot A_0^3 + \frac{3 q_1}{ 4 q_2}\,\,A_0^3\,\, V^{-(q_1 + 4/3)} \dot
V &=& \frac{b}{\kappa} \sqrt{q_2^5 q_3} V^{5q_1/2 - 2/3}.
\label{eqa31}
\end{eqnarray}
\end{subequations}

So finally in view of \eqref{A30Inv} equation \eqref{EqVBIX} $V$ can
be rewritten as

\begin{eqnarray}
\ddot V &-& \frac{3 \kappa}{2} \sqrt{\frac{q_2^3}{q_3}}
\frac{h^\prime}{f^\prime}\dot A_0^3 V^{3q_1/2} - \frac{1}{4} q_2^4
V^{4q_1 + 1/3}\left(\frac{h^\prime}{f}\right)^2 - \frac{1}{2} q_2
q_3 V^{q_1 + 1/3}\frac{f^{\prime\prime}}{f}\nonumber \\ &=& \frac{3
 \kappa}{2}\left[m_{\rm sp} S + 2 \left(F - K F_K\right)\right] V, \label{EqVBIXnew}
\end{eqnarray}

To find the solution to the equation \eqref{EqVBIXnew} we have to
give the concrete form of spinor field nonlinearity. Following some
previous papers, we choose the nonlinearity to be the function of
$S$ only, having the form
\begin{equation}
F = \sum_{k} \lambda_k I^{n_k} =  \sum_{k} \lambda_k S^{2 n_k}.
\label{nonlinearity}
\end{equation}
Recently, this type of nonlinearity was considered in a number of
papers \cite{Saha2015CJP,Saha2015CnJP,Saha2016IJTP,Saha2016EPJP}.

In what follows, we thoroughly study the cases with trivial and
non-tribal $b$.

{\bf Case 1. $b = 0$}

Here we consider the simplest possible case setting $b = 0.$ This
case coincides with the one that corresponds to the diagonal
energy-momentum tensor.

From \eqref{BIX} we have $f = \sin (x_3)$. Solving \eqref{eqh} in
this case one finds $h = \cos(x_3)$.  It should be emphasized that
this form of Bianchi type-$IX$ metric with $f = \sin (x_3)$ and $h =
\cos (x_3)$ is generally considered in literature. In this case for
$A_0^3$ from \eqref{eqa31} we also find
\begin{eqnarray}
\frac{\dot A_0^3}{A_0^3} + \frac{3 q_1}{ 4 q_2} V^{-(q_1 + 4/3)}
\dot V = 0, \label{A03a}
\end{eqnarray}
with the solution
\begin{eqnarray}
A_0^3 = C_1 \exp{\left[\frac{9q_1}{4q_2 (3q_1 +1)} V^{-(q_1 +
1/3)}\right]}. \label{A03}
\end{eqnarray}

Now  on account of $S = V_0/V$ equation \eqref{EqVBIXnew} together
with \eqref{eqa31} can be written as

\begin{subequations}
\label{YVA30}
\begin{eqnarray}
\dot V & = & Y, \label{Vfin0}\\
\dot Y & = & -\frac{3 \kappa}{2} \sqrt{q_2^3/q_3} \tan(x_3) V^{3q_1/2} \Phi_1(V, A_0^3, Y) + \Phi_2(V, A_0^3, Y), \label{Yfin0}\\
\dot A_0^3 &=& \Phi_1(V, A_0^3, Y). \label{A03fin0}
\end{eqnarray}
\end{subequations}

\begin{eqnarray}
\Phi_1 (V, A_0^3, Y) &=& -\frac{3 q_1}{ 4 q_2}\,\,A_0^3\,\, V^{-(q_1
+ 4/3)} Y,\nonumber\\
\Phi_2(V, A_0^3, Y) &=&  \frac{1}{4} q_2^4 V^{4q_1 + 1/3}
\left(\frac{h^\prime}{f}\right)^2 +
\frac{1}{2} q_2 q_3 V^{q_1 + 1/3} \frac{f^{\prime\prime}}{f} \nonumber\\
 &+&   \frac{3}{2}
\left[\left(m_{\rm sp} + \lambda_0\right) + 2 \lambda_1( 1 - n_1)
V^{1 - 2n_1} + 2 \lambda_2( 1 - n_2) V^{1 - 2n_2} \right].\nonumber
\end{eqnarray}

In what follows, we solve the foregoing equation numerically. For
this reason we first rewrite it as a system of equations in the
following way:

For simplicity we consider only three terms of the sum. We set $n_k
= n_0: 1 - 2 n_0 = 0$ which gives $n_0 = 1/2$. In this case the
corresponding term can be added with the mass term. We assume that
$q_1$ is a positive quantity, so that $4q_1 + 1/3$ is positive too.
For the nonlinear term to be dominant at large time, we set $n_k =
n_1: 1 -2 n_1 > 4q_1 + 1/3$, i.e., $n_1 < 1/3 - 2 q_1 $. And
finally, for the nonlinear tern to be dominant at the early stage we
set $n_k = n_2: 1 - 2 n_2 < 0$, i.e., $n_2 > 1/2$. Since we are
interested in qualitative picture of evolution, let us set $q_2 =
1,\, q_3 = 1$ and $\kappa = 1$. We also assume $V_0 = 1$. Then we
have

\begin{eqnarray}
\Phi_2(V, A_0^3, Y) &=&  \frac{1}{4}  V^{4q_1 + 1/3} -
 \frac{1}{2}  V^{q_1 + 1/3}  \nonumber\\
 &+&   \frac{3}{2}
\left[\left(m_{\rm sp} + \lambda_0\right) + 2 \lambda_1( 1 - n_1)
V^{1 - 2n_1} + 2 \lambda_2( 1 - n_2) V^{1 - 2n_2} \right],\nonumber
\end{eqnarray}
We set $m_{\rm sp} = 1$ and $l_0 = 2$. As far as $q_1$, $n_1$ and
$n_2$ are concerned, in line of our previous discussions we choose
them in such a way that the power of nonlinear term in the equations
become integer. We have also studied the case for some different
values, but they didn't give any principally different picture. We
choose $q_1 = 2/3$,\, $n_1 = -3/2 < 1/3 - 2 q-1 = -1$ and $n_2 = 3/2
> 1/2$. It should be noted that we have taken some others value for
$q_1$ such as $q_1 = -1$, but it does not give qualitatively
different picture. We have also set $x_3 = [0,\,\pi] = k \pi/5 $
with step $k = 0..5$. Finally we have considered time span $[0,\,2]$
with step size $0.001$. Here we consider different values of
$\lambda_i$ both positive and negative. We choose the initial values
for $V(0) = 1$, $ Y(0) = \dot V(0) = 0.1$, and $A_0^3 (0) = 1$,
respectively.

\begin{figure}[ht]
\centering
\includegraphics[width=75mm]{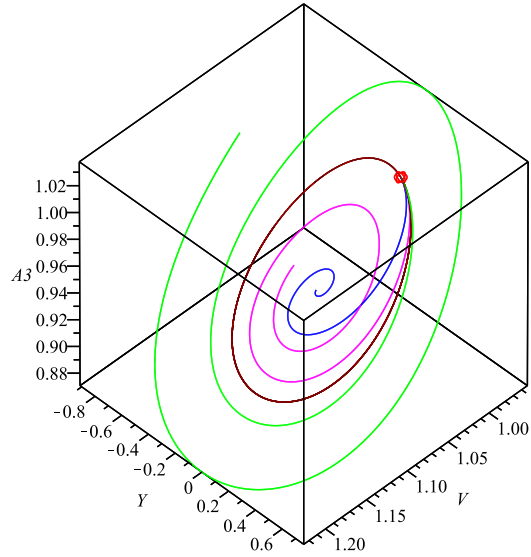}\\ \vskip 2 cm
\caption{Phase diagram of $[V,\, \dot V,\, A_0^3]$ in case of
$\lambda_1 = 1$ and $\lambda_2 = 1$.} \label{BIX-3Db0l2p}.
\end{figure}

\begin{figure}[ht]
\centering
\includegraphics[width=75mm]{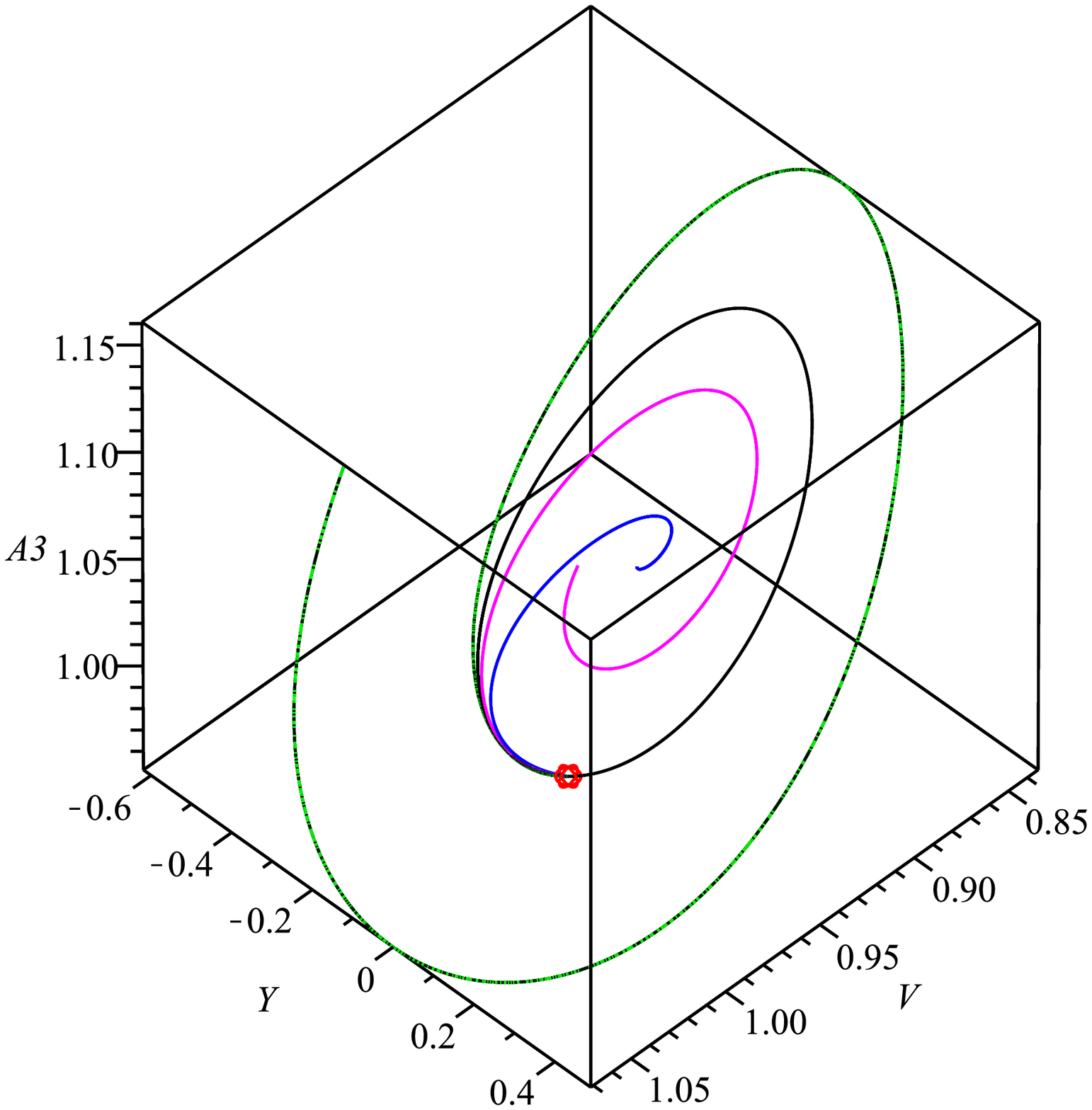}\\ \vskip 2 cm
\caption{Phase diagram of $[V,\, \dot V,\, A_0^3]$ in case of
$\lambda_1 = 1$ and $\lambda_2 = -0.1$.} \label{BIX-3Db0l2m}.
\end{figure}

In  Figs. \ref{BIX-3Db0l2p} and \ref{BIX-3Db0l2m} we have plotted
the phase diagram of $[V,\, \dot V,\, A_0^3]$ for both positive and
negative $\lambda_2$, respectively. In both cases $\lambda_1 = 1$.
Analogical picture was found for $\lambda_2 =0$ and $\lambda_1 = 1$.

\begin{figure}[ht]
\centering
\includegraphics[width=75mm]{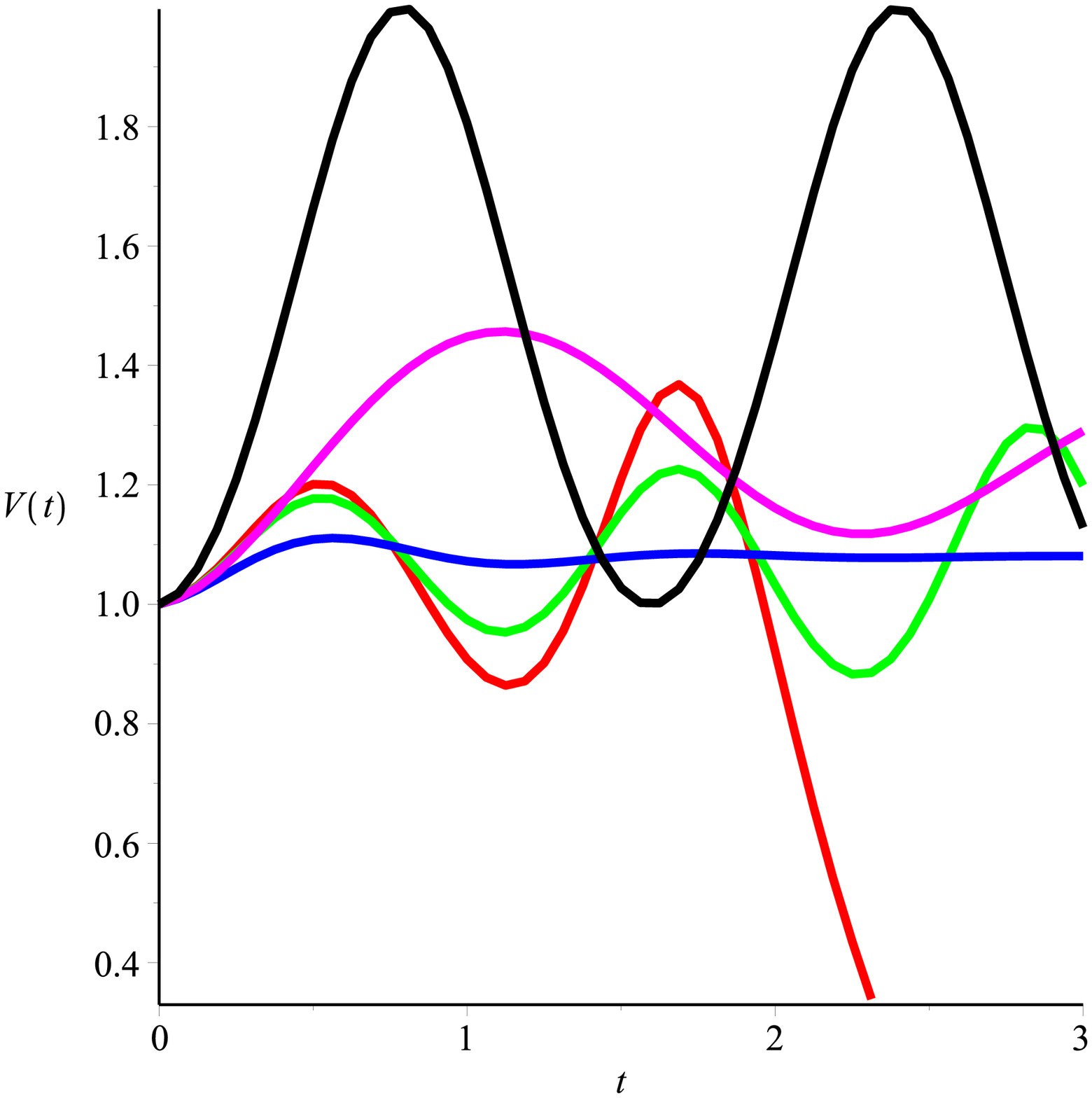}\\ \vskip 1 cm
\caption{Evolution of $V$ in case of $\lambda_1 = 1$ and $\lambda_2
= 1$.} \label{BIX-Vb0l2p}.
\end{figure}

\begin{figure}[ht]
\centering
\includegraphics[width=75mm]{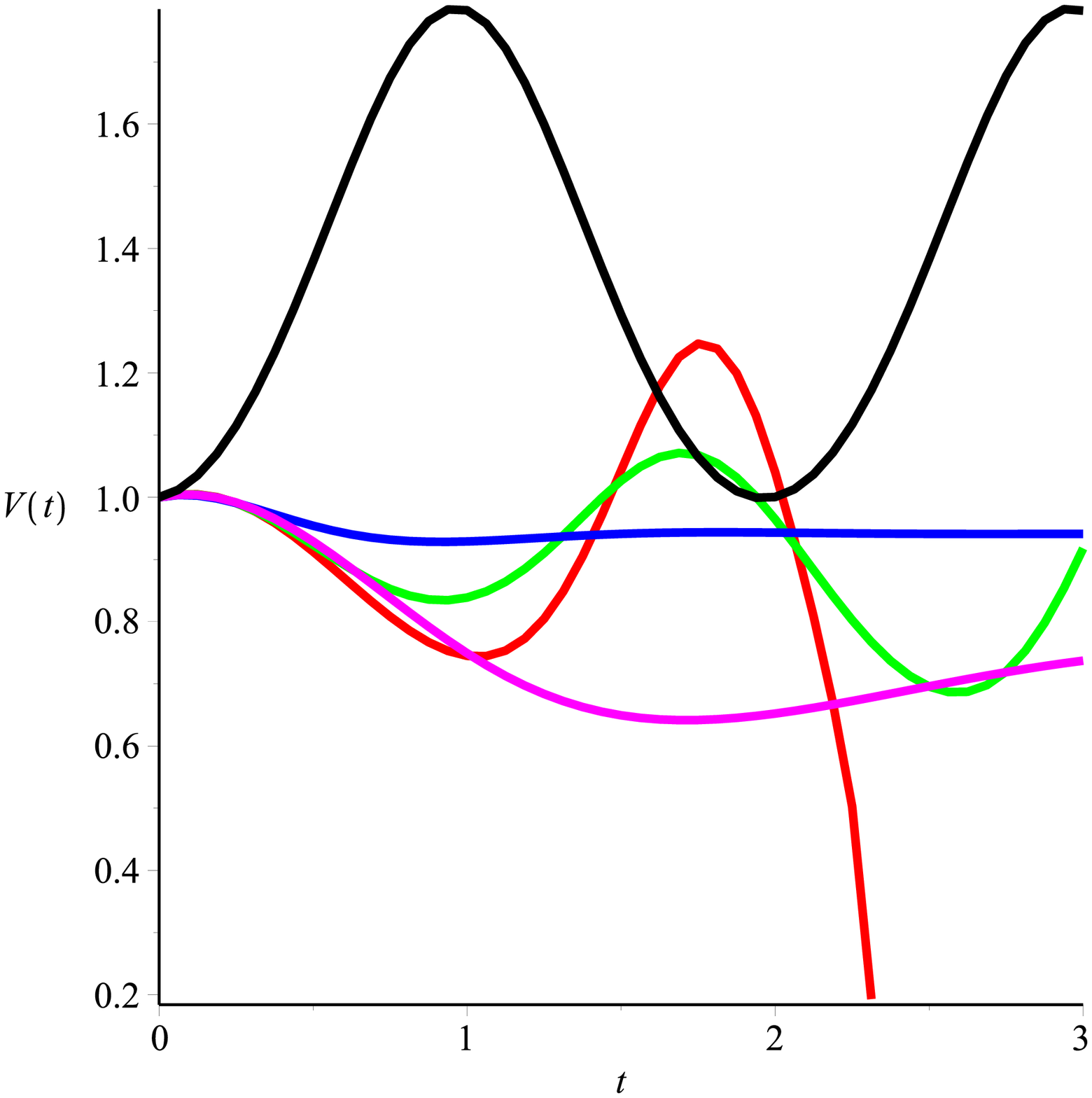}\\ \vskip 1 cm
\caption{Evolution of $V$ in case of $\lambda_1 = 1$ and $\lambda_2
= -0.1$.} \label{BIX-Vb0l2m}.
\end{figure}

In Figs. \ref{BIX-Vb0l2p} and \ref{BIX-Vb0l2m} evolution of $V$
corresponding to Figs. \ref{BIX-3Db0l2p} and \ref{BIX-3Db0l2m} are
demonstrated. As one sees, in both cases we have oscillatory mode of
expansion.

\begin{figure}[ht]
\centering
\includegraphics[width=75mm]{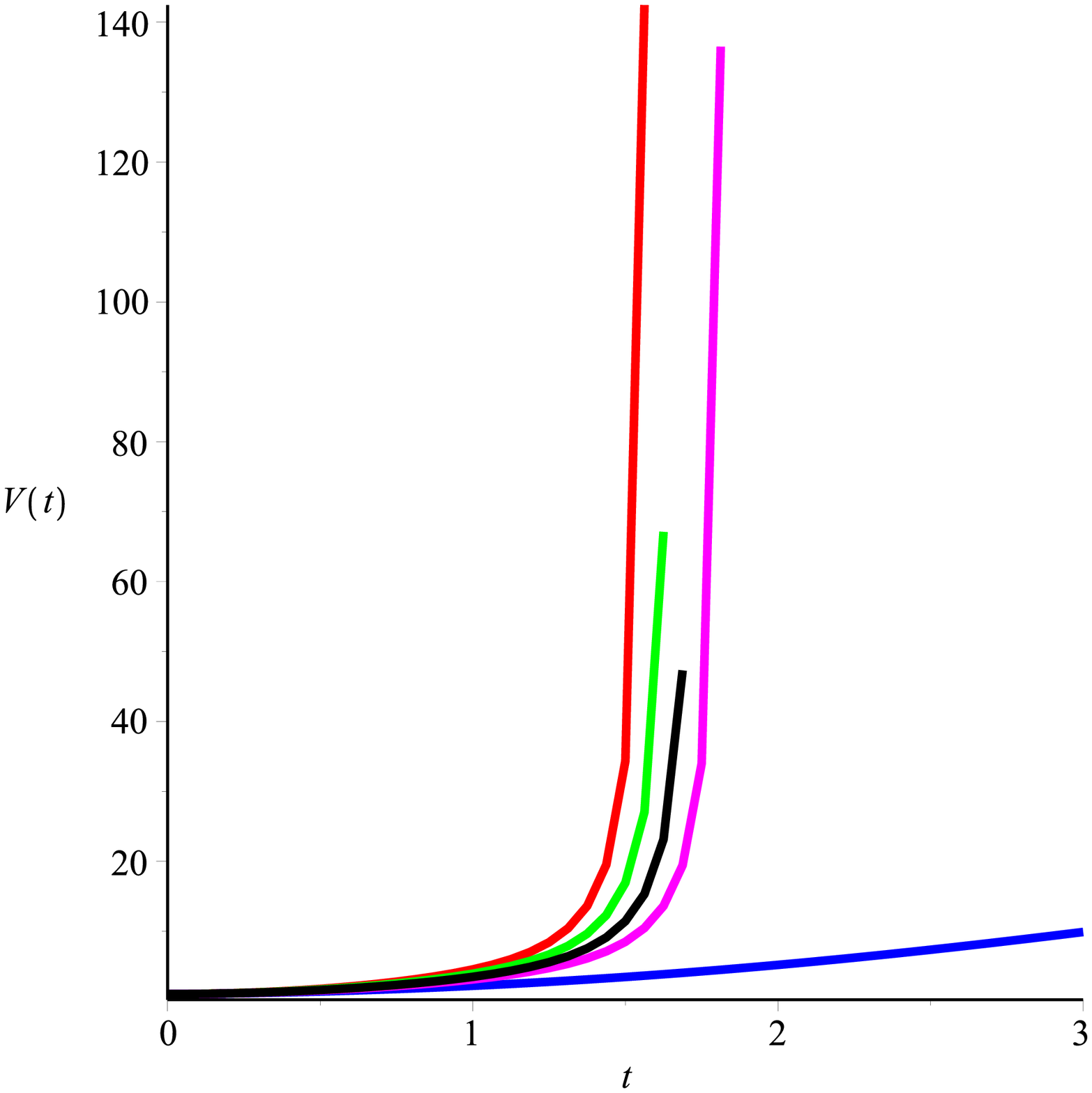}\\ \vskip 1 cm
\caption{Evolution of $V$ in case of $\lambda_1 = 0$ and $\lambda_2
= 0$.} \label{BIX-Vb0l20l10}.
\end{figure}

\begin{figure}[ht]
\centering
\includegraphics[width=75mm]{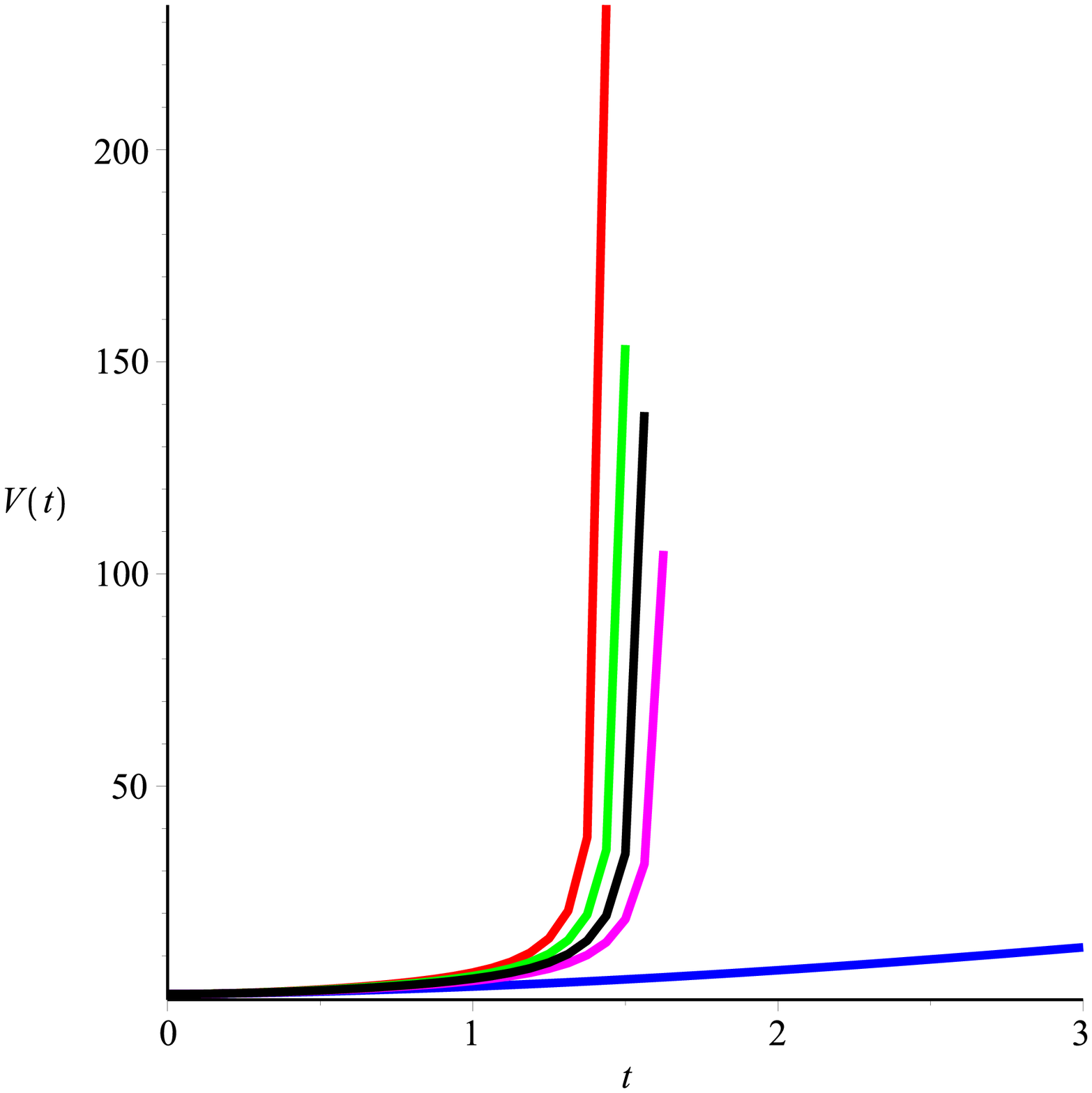}\\ \vskip 1 cm
\caption{Evolution of $V$ in case of $\lambda_1 = 0$ and $\lambda_2
= 1$.} \label{BIX-Vb0l21l10}.
\end{figure}

In Figs. \ref{BIX-Vb0l20l10} and \ref{BIX-Vb0l21l10} we have
illustrated the evolution of $V$ for $\lambda_1 = 0$ and $\lambda_2
= 0$ (linear spinor field) and for $\lambda_1 = 0$ and $\lambda_2 =
1$, respectively. This shows that in case of $\lambda_1 = 0$ we have
a rapid expansion of $V$ at a very early stage.

In the Figures each color corresponds to a concrete value of $x_3 =
k \pi/5$, namely, red, green, yellow, blue, magenta and black color
corresponds to $k = 0,1,2,3,4,5$

{\bf Case 2. $b \ne 0$}

Let us consider the case with non-trivial $b$. Inserting $f = \sin
(x_3)$ into \eqref{eqh} one finds

\begin{equation}
h(x_3) = b\left(\sin(x_3) - x_3 \cos(x_3)\right) - c_1 \cos(x_3) +
c_2, \quad c_1 = {\rm const.}, \quad c_2 = {\rm const.} \label{hb}
\end{equation}

The equations  \eqref{EqVBIXnew} together with \eqref{eqa31} in this
case can be written as

\begin{subequations}
\label{YVA3}
\begin{eqnarray}
\dot V & = & Y, \label{Vfin}\\
\dot Y & = & \frac{3 \kappa}{2} \sqrt{\frac{q_2^3}{q_3}} \frac{h^\prime}{f^\prime}V^{3q_1/2} \Phi_1(V, A_0^3, Y) +
\Phi_2(V, A_0^3, Y), \label{Yfin}\\
\dot A_0^3 &=& \Phi_1(V, A_0^3, Y). \label{A03fin}
\end{eqnarray}
\end{subequations}

where we denote

\begin{eqnarray}
\Phi_1 (V, A_0^3, Y) &=& -\frac{3 q_1}{ 4 q_2}\,\,A_0^3\,\, V^{-(q_1
+ 4/3)} Y
+ \frac{b}{\kappa} \sqrt{q_2^5 q_3} V^{5q_1/2 - 2/3},\nonumber\\
\Phi_2(V, A_0^3, Y) &=&  \frac{1}{4} q_2^4 V^{4q_1 + 1/3}
\left(\frac{h^\prime}{f}\right)^2 +
\frac{1}{2} q_2 q_3 V^{q_1 + 1/3} \frac{f^{\prime\prime}}{f} \nonumber\\
 &+&  \frac{3\kappa}{2}
\left[\left(m_{\rm sp} + \lambda_0\right) + 2 \lambda_1( 1 - n_1)
V^{1 - 2n_1} + 2 \lambda_2( 1 - n_2) V^{1 - 2n_2} \right].\nonumber
\end{eqnarray}

In what follows, we solve the foregoing system numerically for $b =
1$ with all other parameters taking same value as corresponding
cases for a trivial $b$.

\begin{figure}[ht]
\centering
\includegraphics[width=75mm]{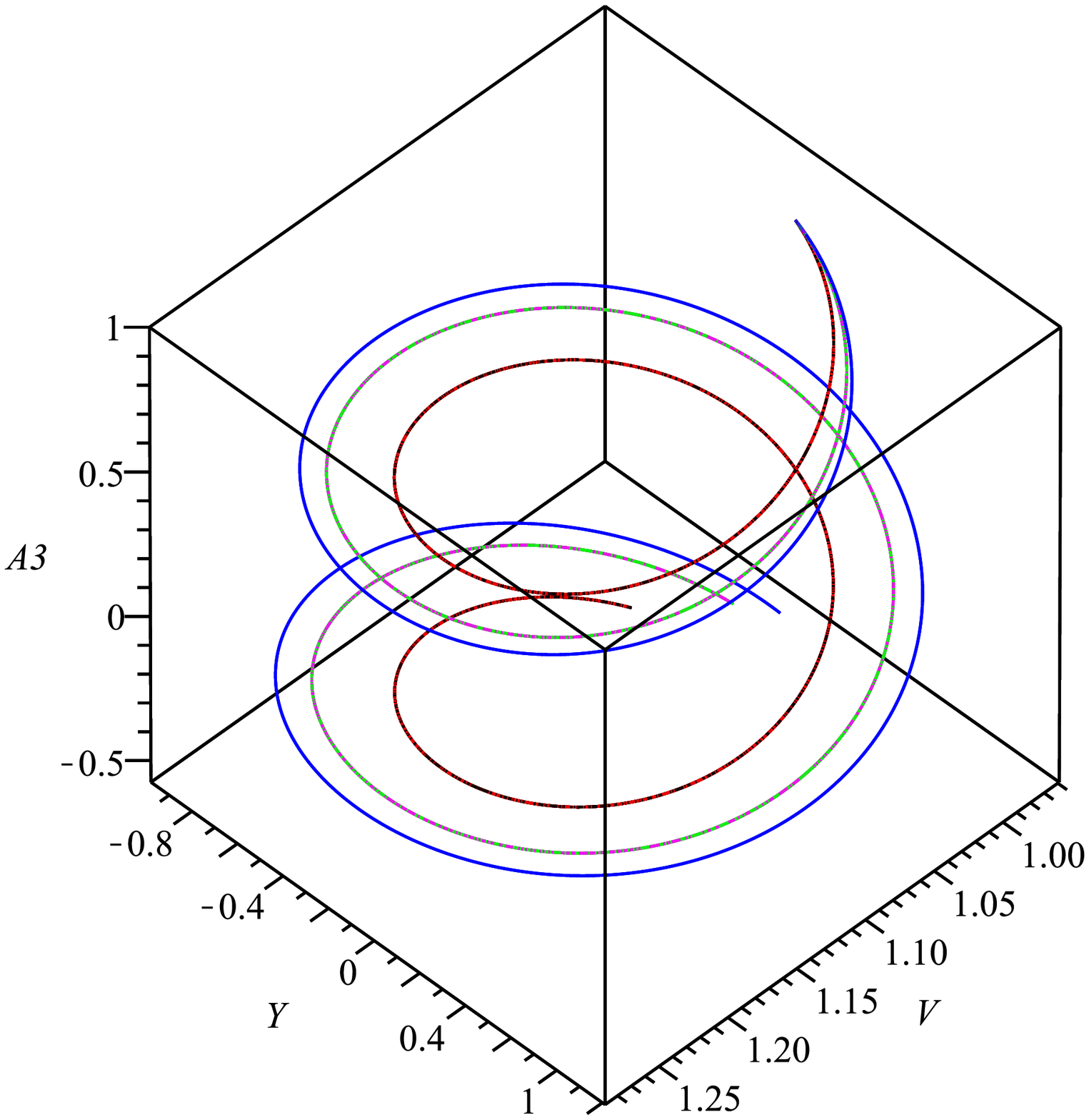}\\ \vskip 2 cm
\caption{Phase diagram of $[V,\, \dot V,\, A_0^3]$ in case of
$\lambda_1 = 1$ and $\lambda_2 = 1$.} \label{BIX-3Db1l2p}.
\end{figure}

\begin{figure}[ht]
\centering
\includegraphics[width=75mm]{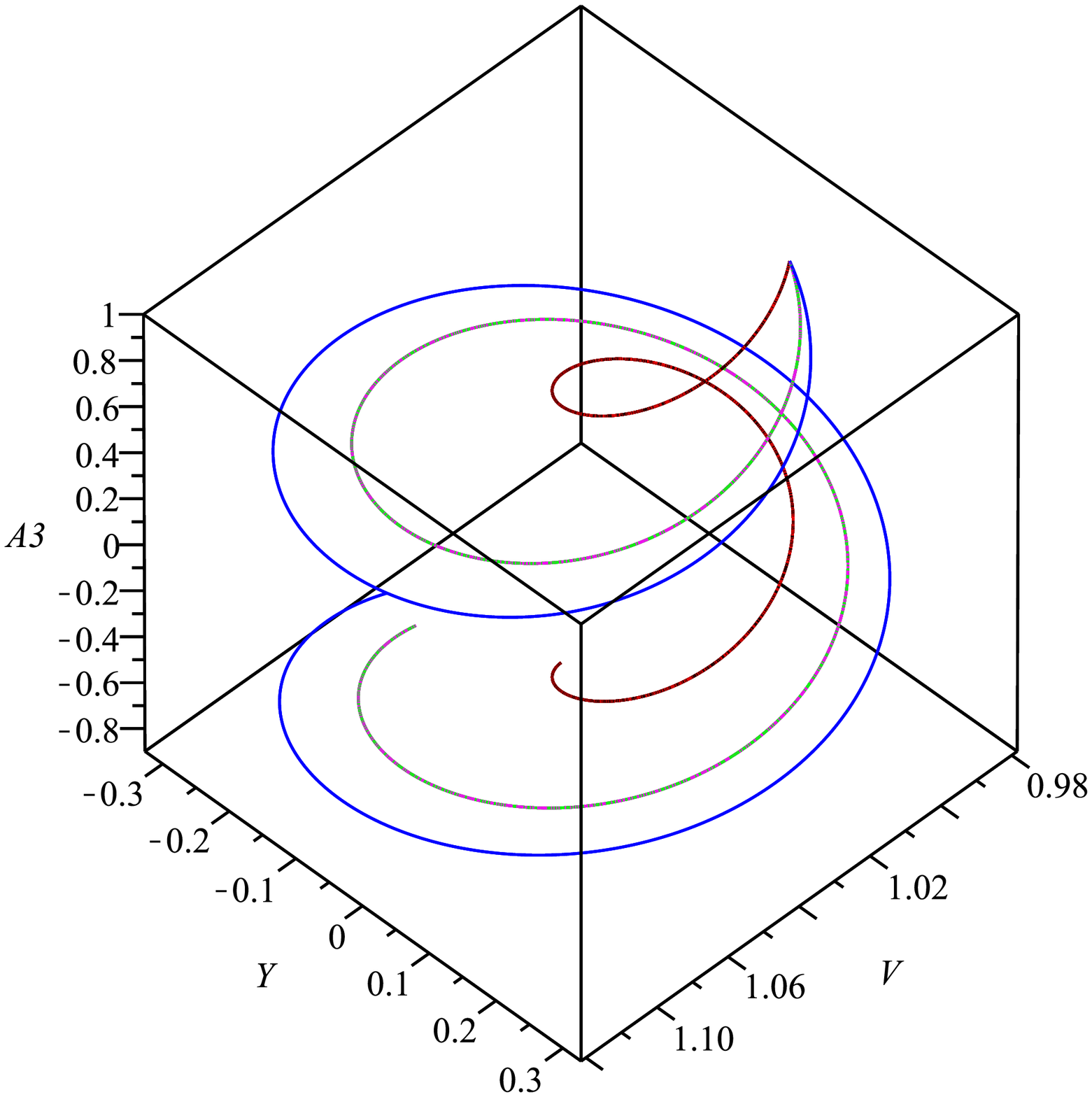}\\ \vskip 2 cm
\caption{Phase diagram of $[V,\, \dot V,\, A_0^3]$ in case of
$\lambda_1 = 1$ and $\lambda_2 = -0.1$.} \label{BIX-3Db1l2m}.
\end{figure}

In  Figs. \ref{BIX-3Db1l2p} and \ref{BIX-3Db1l2m} we have plotted
the phase diagram of $[V,\, \dot V,\, A_0^3]$ for both positive and
negative $\lambda_2$, respectively. In both cases $\lambda_1 = 1$.
Analogical picture was found for $\lambda_2 =0$ and $\lambda_1 = 1$.
As one sees, while in case of a trivial $b$ we have a focus like
phase diagram, in this case with non-zero $b$ the phase diagram is a
spiral.

\begin{figure}[ht]
\centering
\includegraphics[width=75mm]{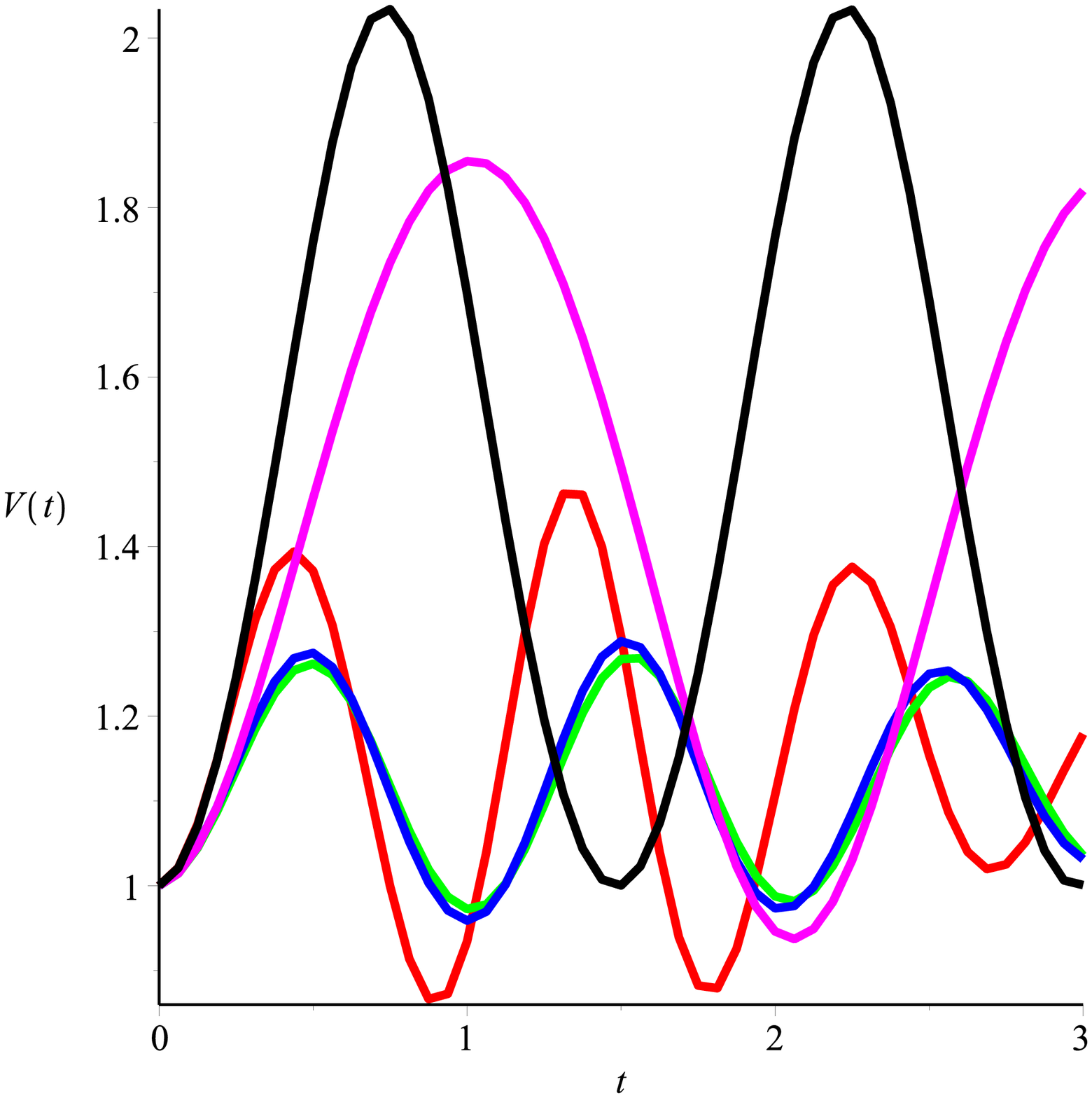}\\ \vskip 1 cm
\caption{Evolution of $V$ in case of $\lambda_1 = 1$ and $\lambda_2
= 1$.} \label{BIX-Vb1l2p}.
\end{figure}

\begin{figure}[ht]
\centering
\includegraphics[width=75mm]{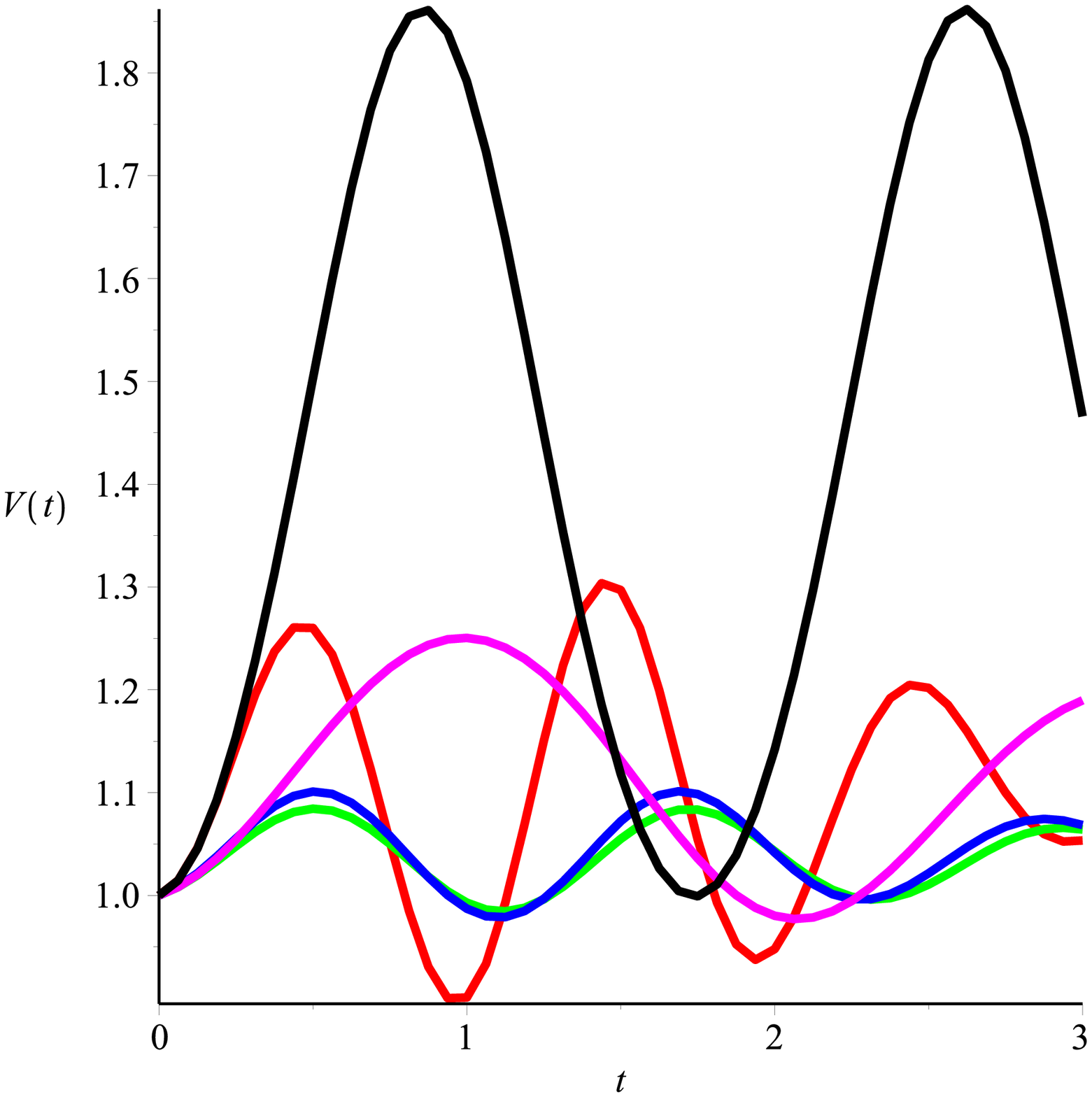}\\ \vskip 1 cm
\caption{Evolution of $V$ in case of $\lambda_1 = 1$ and $\lambda_2
= -0.1$.} \label{BIX-Vb1l2m}.
\end{figure}

In Figs. \ref{BIX-Vb1l2p} and \ref{BIX-Vb1l2m} evolution of $V$
corresponding Figs. \ref{BIX-3Db1l2p} and \ref{BIX-3Db1l2m} are
demonstrated. As one sees, in both cases we have oscillatory mode of
expansion.

\begin{figure}[ht]
\centering
\includegraphics[width=75mm]{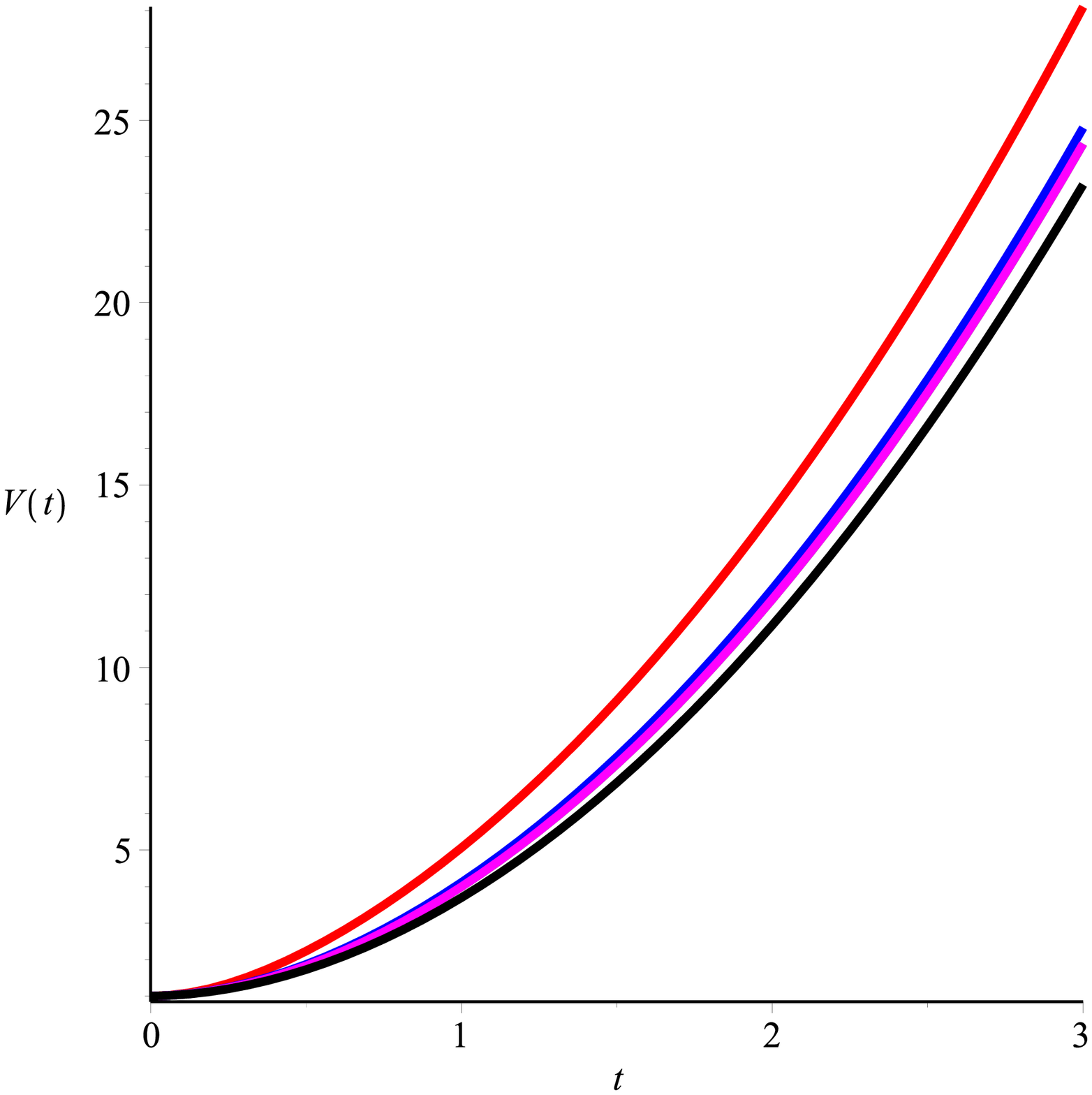}\\ \vskip 1 cm
\caption{Evolution of $V$ in case of $\lambda_1 = 0$ and $\lambda_2
= 0$.} \label{BIX-Vb1l20l10}.
\end{figure}

\begin{figure}[ht]
\centering
\includegraphics[width=75mm]{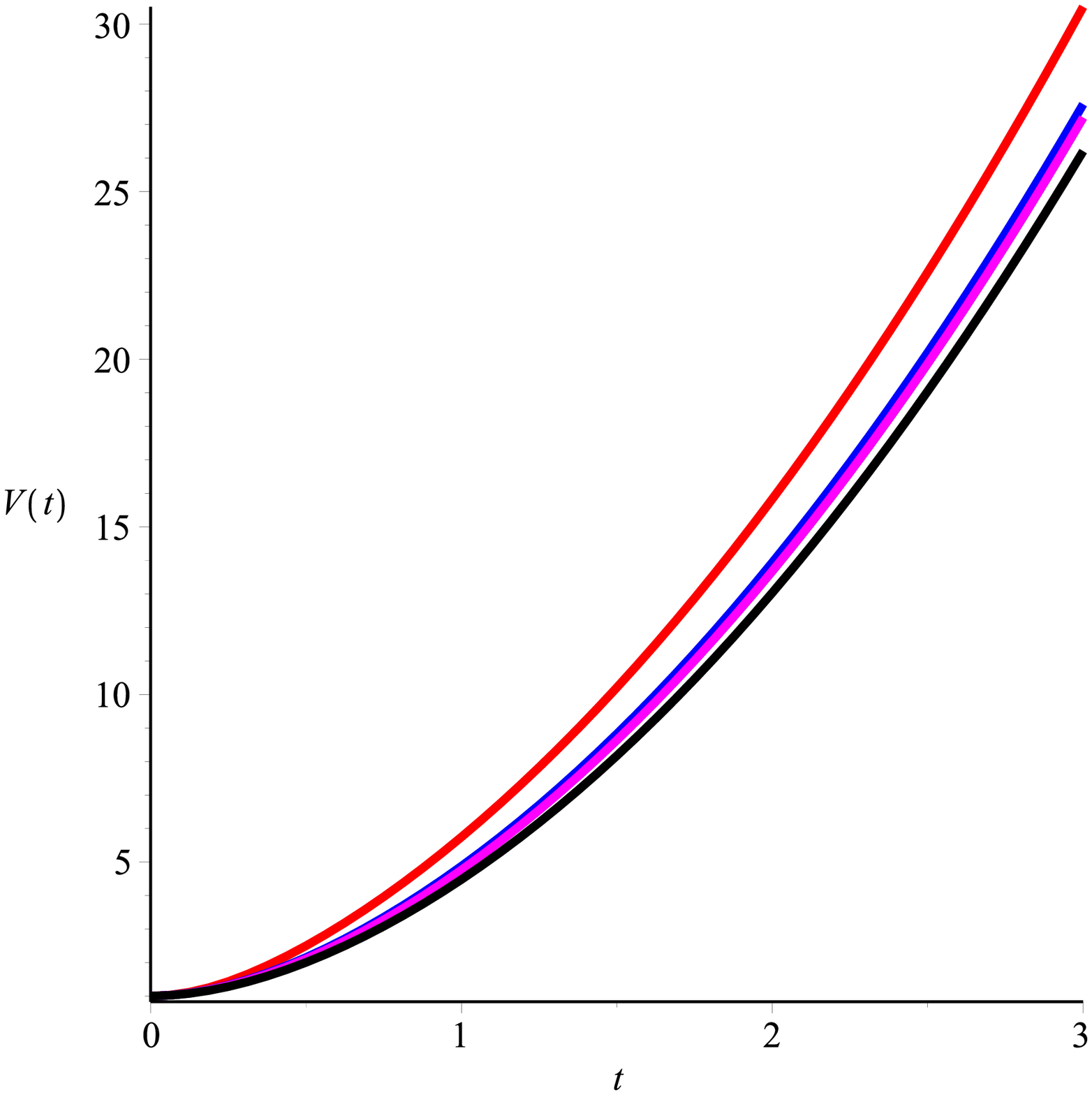}\\ \vskip 1 cm
\caption{Evolution of $V$ in case of $\lambda_1 = 0$ and $\lambda_2
= 1$.} \label{BIX-Vb1l21l10}.
\end{figure}

In Figs. \ref{BIX-Vb1l20l10} and \ref{BIX-Vb1l21l10} we have
illustrated the evolution of $V$ for $\lambda_1 = 0$ and $\lambda_2
= 0$ (linear spinor field) and for $\lambda_1 = 0$ and $\lambda_2 =
1$, respectively. This shows that in case of $\lambda_1 = 0$ we have
a rapid expansion of $V$ at a very early stage.

\section{Conclusion}

Within the scope of Bianchi type-$IX$ cosmological model we have
studied the role of spinor field in the evolution of the Universe.
It is found that unlike the diagonal Bianchi models in this case the
components of energy-momentum tensor of the spinor field along the
principal axis are not the same, i.e. $T_1^1 \ne T_2^2 \ne T_3^3$,
even in absence of spinor field nonlinearity. The presence of
nontrivial non-diagonal components of energy-momentum tensor of the
spinor field imposes severe restrictions both on geometry of
space-time and on the spinor field itself. As a result the
space-time turns out to be either locally rotationally symmetric or
isotropic. In this paper we considered the Bianchi type-$IX$
space-time both for a trivial $b$, that corresponds to standard
$BIX$ and the one with a non-trivial $b$.  It was found that a
positive $\lambda_1$ gives rise to an oscillatory mode of expansion,
while a trivial $\lambda_1$ leads to rapid expansion at the early
stage of evolution.

\vskip 0.1 cm

\noindent {\bf Acknowledgments}\\
This work is supported in part by a joint Romanian-LIT, JINR, Dubna
Research Project, theme no. 05-6-1119-2014/2016.

I would also like to thank Victor Rikhvitsky for constant help in
numerical solutions.

\end{document}